# Spinor-Induced Instability of Kinks, Holes and Quantum Droplets

Yaroslav V. Kartashov[1,2], V. M. Lashkin[3,4], Michele Modugno[5,6], and Lluis Torner[1,7]

[1]*ICFO-Institut de Ciencies Fotoniques, The Barcelona Institute of Science and Technology, 08860 Castelldefels (Barcelona), Spain*
[2]*Institute of Spectroscopy, Russian Academy of Sciences, Troitsk, Moscow, 108840, Russia*
[3]*Institute for Nuclear Research, Pr. Nauki 47, Kyiv 03028, Ukraine*
[4]*Space Research Institute, Pr. Glushkova 40 k.4/1, Kyiv 03187, Ukraine*
[5]*Department of Physics, University of the Basque Country UPV/EHU, 48080 Bilbao, Spain*
[6]*IKERBASQUE Basque Foundation for Science, 48013 Bilbao, Spain*
[7]*Universitat Politecnica de Catalunya, 08034 Barcelona, Spain*

We address the existence and stability of one-dimensional (1D) holes and kinks and two-dimensional (2D) vortex-holes nested in extended binary Bose mixtures, which emerge in the presence of Lee-Huang-Yang (LHY) quantum corrections to the mean-field energy, along with self-bound quantum droplets. We consider both the symmetric system with equal intra-species scattering lengths and atomic masses, modeled by a single (scalar) LHY-corrected Gross-Pitaevskii equation (GPE), and the general asymmetric case with different intra-species scattering lengths, described by two coupled (spinor) GPEs. We found that in the symmetric setting, 1D and 2D holes can exist in a stable form within a range of chemical potentials that overlaps with that of self-bound quantum droplets, but that extends far beyond it. In this case, holes are found to be always stable in 1D and they transform into pairs of stable out-of-phase kinks at the critical chemical potential at which localized droplets turn into flat-top states, thereby revealing the connection between localized and extended nonlinear states. In contrast, we found that the spinor nature of the asymmetric systems may lead to instability of 1D holes, which tend to break into two gray states moving in the opposite directions. Remarkably, such instability arises due to spinor nature of the system and it affects only holes nested in extended modulationally-stable backgrounds, while localized quantum droplet families remain completely stable, even in the asymmetric case, while 1D holes remain stable only close to the point where they transform into pairs of kinks. We also found that symmetric systems allow fully stable 2D vortex-carrying single-charge states at moderate amplitudes, while unconventional instabilities appear also at high amplitudes. Symmetry also strongly inhibits instabilities for double-charge vortex-holes, which thus exhibit unexpectedly robust evolutions at low amplitudes.

The behavior of nonlinear wave excitations and their dynamical stability and interaction properties in systems characterized by the presence of competing nonlinearities may dramatically differ from those encountered in systems governed by the non-linear Schrödinger and Gross-Pitaevskii equations featuring only a cubic nonlinearity. Competition between nonlinearities of different orders or physical nature may lead to stabilization of otherwise unstable fundamental or excited states, it causes unusual shape transformations and, most importantly, it may lead to the existence or merger of nonlinear states that otherwise would not simultaneously appear in the system, or even to the emergence of previously unknown self-sustained states. Such phenomena are known to occur in a broad range of physical systems, including optical, optoelectronic, acoustic and matter waves [1-3]. In this content, it was discovered in Bose-Einstein condensates that the Lee-Huang-Yang (LHY) correction to the mean-field energy due to quantum fluctuations [4] that becomes important when other nonlinear interactions counterbalance each other, may suppress collapse and thus lead to the formation of stable multidimensional quantum droplets [5,6]. The LHY quantum corrections manifest themselves as competing nonlinearities in the modified Gross-Pitaevskii equation (GPE) governing the system. Importantly, as shown for Bose-Bose mixtures [5,6], the mathematical form of the LHY correction and, consequently, the ensuing behavior of nonlinear states depend dramatically on the dimensionality of the system (for recent reviews, see [7,8]).

Quantum droplets have been observed experimentally and analyzed theoretically in 2D and 3D geometries in a single-component dipolar Bose gas, where collapse driven by long-range attractive dipolar interactions can be compensated by the repulsive LHY contribution [9-18] and in Bose-Bose mixtures [19-26], where the repulsive LHY correction becomes important when the intra-component repulsion is compensated by the inter-component attraction. Their static and dynamical properties have also been investigated in 1D systems [27-32] where, remarkably, the role of the mean-field and LHY energy terms are reversed, the latter providing the attractive contribution needed to form bound states [6,27]. The collisions of droplets have been also addressed experimentally [22] and theoretically [28,33]. In higher-dimensional settings, it was found that the LHY contribution can suppress azimuthal instabilities and stabilize localized vortex droplets in 2D [34,35] and 3D [36] Bose-Bose mixtures, and can lead to the formation of robust dynamical states, such as rotating droplets and droplet clusters [37-40]. Modulational instabilities that may produce sets of droplets was analyzed in [41,42] for dipolar condensates. Also, stable 2D and 3D quantum droplets have been studied in dipolar mixtures [42,43], and their properties may be significantly affected by optical lattices [44,45] (for the corresponding one-dimensional case see Refs. [46,47]) and spin-orbit interactions [48-50]. In dipolar droplets particular attention has been devoted to the formation of so-called supersolid behavior (see [51,52] and the review [8]). So-called LHY fluids in the regime with vanishing cubic nonlinearity, whose dynamics is governed by quantum fluctuations [53], have been observed recently [54]. The pairing theory for binary Bose mixtures with interspecies attractions elucidating the regimes of formation of multidimensional quantum droplets was developed in [55].

Most previous efforts have been devoted to the study of self-bound quantum droplets, where the LHY correction serves as stabilizing mechanism to suppress collapse or the instabilities of excited states. However, it has not been yet widely appreciated that it may be possible to embed different types of localized features in the LHY dominated regime, thus generating new nonlinear global states, such as the recently predicted quantum bubbles arising with unequal intraspecies interactions or unequal masses [56]. Holes nested in extended droplets whose background may be made modulationally stable in different dimensionalities are another potential fascinating possibility, which to the date has not been explored.

The goal of this paper is twofold. First, we aim at showing that in the regimes in which LHY quantum corrections are relevant, Bose-Bose mixtures support stable 1D kinks (i.e. nonlinear states with different asymptotic densities along the two transverse directions) and holes (states with identical asymptotic densities, but with a phase jump in the center where a dark notch develops in the density profile), as well as 2D vortex-holes (vortices with different topological charges nested in an extended background), and that the LHY correction substantially impacts the properties of such states, especially at low densities. Second, we compare the properties and stability of kinks and holes in the simplest symmetric model with equal atomic masses and intra-species interactions, described by a single GPE derived in [5, 6], with those of the more general asymmetric model involving two coupled GPEs with different intra-species interactions. Note that the single wave equation model was employed to predict self-bound quantum droplets in binary Bose mixtures [5], and it represents an interesting theoretical model in its own right. Indeed, it was later used for the description of quantum droplets in different configurations and dimensionalities (see, e.g., Refs. [6,28,46]), as well as for the description of the LHY fluid [53,54,57]. It is useful to elucidate also the properties of the asymmetric model when the single wave function ansatz is released and each component is described by its own wavefunction.

We found that both models predict that 1D hole solutions can be found in addition to self-bound droplets and that they transform into a pair of kinks at the critical value of the chemical potential at which self-bound quantum droplets cease to exist. Actually, when the chemical potential approaches this value self-bound droplets acquire flat-top shapes, while their width greatly increases at nearly constant maximal density. This reveals a connection between localized and extended states that is unique for this system and is not observed for conventional dark and vortex matter-wave solitons [58-68]. Remarkably, we also found that the stability properties of 1D holes are dramatically different in the symmetric and asymmetric systems: While in the former 1D holes are always stable in their entire range of existence, according to the spinor asymmetric model they become unstable, breaking up into two gray states moving in the opposite directions. This instability is unique, as it occurs only for extended states, while localized quantum droplets remain stable. As a matter of fact, 1D holes in the asymmetric model are stable only close to the point of their transformation into a pair of kinks. In 2D systems, the logarithmic nonlinearity induced by the LHY effects allows stable single-charge vortex-holes at moderate background amplitudes, and leads to unconventional instabilities at high amplitudes. It also strongly inhibits instabilities for double-charge vortex-holes that are usually unstable without external traps [69-73].

We start with the simplest 1D model for a Bose-Bose mixture with equal masses and intra-species interactions, under the assumption that the two components can be described by a single wave function, namely $\psi_1 = \psi_2 = \psi$. The corresponding modified Gross-Pitaevskii equation takes the following form (in dimensionless units):

$$i\frac{\partial \psi}{\partial t} = -\frac{1}{2}\frac{\partial^2 \psi}{\partial x^2} + \delta g |\psi|^2 \psi - g_{\mathrm{LHY}} |\psi| \psi. \quad (1)$$

This model, derived in the seminal work [6], has been successfully used for the prediction of 1D self-bound quantum droplets. Here $x = X r_0^{-1}$ is the scaled coordinate, $r_0$ is the characteristic spatial scale defining the characteristic energy $\varepsilon_0 = \hbar^2 / m r_0^2$ and time $t_0 = \hbar \varepsilon_0^{-1}$ scales ($t = T t_0^{-1}$), $m = m_1 = m_2$ is the atomic mass of the two components, $\delta g = g_{12} + (g_{11}g_{22})^{1/2}$ is the effective cubic nonlinearity coefficient determined by the one-dimensional intra-species ($g_{11} = g_{22} > 0$) and inter-species ($g_{12} < 0$) coupling constants (for the relation between the 1D coupling constants and the corresponding scattering lengths see, e.g., Ref. [74]). Here we work in the regime $\delta g > 0$, where effective cubic nonlinearity in Eq. (1) is repulsive, and we assume that it competes with quadratic LHY contribution, whose strength in symmetric case ($g_{11} = g_{22}$) is defined by the coefficient $g_{\mathrm{LHY}} = 2^{1/2} g_{11}^{3/2} / \pi > 0$ [6]. Generalization of this model to the case of unequal components will be considered below. A comparison with the Monte-Carlo simulations for one-dimensional Bose mixtures performed in [74] indicates that the Gross-Pitaevskii equation (1) provides a sufficiently accurate description of the quantum droplet formation for $1.0 > |g_{12}| / (g_{11}g_{22})^{1/2} \gtrsim 0.6$, i.e. for sufficiently small $\delta g$ values. Thus, all results presented here correspond to the regime where this condition is satisfied.

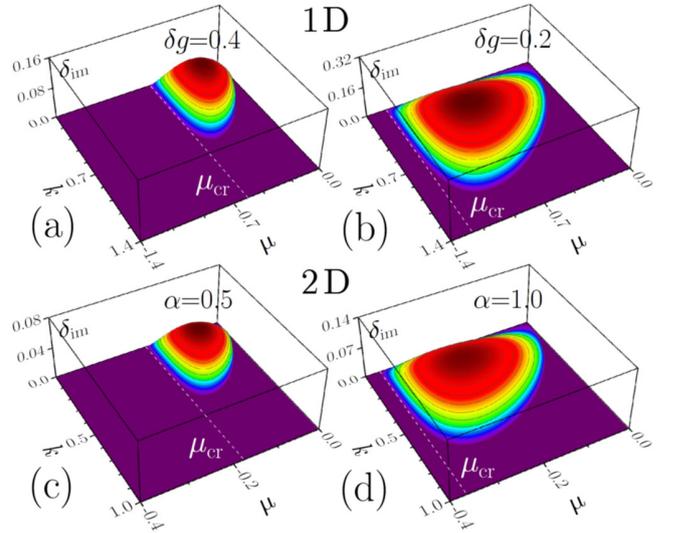

Fig. 1. MI gain spectra for the lower branches of the constant-amplitude solutions in the 1D (a), (b) and 2D (c), (d) symmetric systems. The dashed white lines indicate the critical values of the chemical potential $\mu_{\mathrm{cr}}$. In the 1D case $k = k_{\mathrm{x}}$, while in the 2D case $k = (k_{\mathrm{x}}^2 + k_{\mathrm{y}}^2)^{1/2}$.

First, we consider constant-amplitude solutions of Eq. (1) and their modulational instability (MI). Substitution of constant-amplitude solution $\psi = a e^{-i\mu t}$ into Eq. (1) yields two branches

$$a_{\pm} = \frac{g_{\mathrm{LHY}} \pm (g_{\mathrm{LHY}}^2 + 4\mu \delta g)^{1/2}}{2\delta g} \quad (2)$$

that join at critical value of chemical potential $\mu_{\mathrm{cr}} = -g_{\mathrm{LHY}}^2 / 4\delta g$. The upper branch exists at all $\mu \geq \mu_{\mathrm{cr}}$, while the lower one at $0 \geq \mu \geq \mu_{\mathrm{cr}}$ (its amplitude $a_{-}$ vanishes as $\mu \to 0$). Notice that the presence of two branches of solutions at $0 \geq \mu \geq \mu_{\mathrm{cr}}$ is a specific feature of this system and that it is possible only due to the presence of LHY correction $\sim g_{\mathrm{LHY}}$. Assuming periodic perturbation $\psi = (a_{\pm} + u e^{ikx - i\delta t} + v^* e^{-ikx + i\delta^* t}) e^{-i\mu t}$, where $u, v$ are perturbation amplitudes, $k$ is the modulation wavenumber, one gets:

$$\delta_{\pm} = \frac{k}{2}\left[\frac{(g_{\mathrm{LHY}}^2 + 4\mu\delta g) \pm g_{\mathrm{LHY}}(g_{\mathrm{LHY}}^2 + 4\mu\delta g)^{1/2}}{\delta g} + k^2\right]^{1/2} \quad (3)$$

where $\pm$ signs correspond to the upper/lower branches. One can see that $\delta_\pm$ can acquire nonzero imaginary part (indicating on exponential growth of imposed modulations) only for lower branch, while upper branch is always modulationally stable. The dependencies of growth rate $\delta_{\text{im}} = \text{Im}\,\delta$ on $(\mu, k)$ for lower branch are shown in Figs. 1(a),(b) for various $\delta g$ values at $g_{\text{LHY}} = 1$. The region featuring modulational instability has a finite bandwidth in $k$, which shrinks as $\mu \to \mu_{\text{cr}}$ or $\mu \to 0$, and it broadens with a decrease of $\delta g$ at a fixed $g_{\text{LHY}} = 1$.

Stability of the upper constant-amplitude branch implies that one can nest in such wave stable "holes" – analogs of dark matter-wave solitons [58], modified by LHY correction. The properties of holes are summarized in Figs. 2(a) and 2(b), while examples of their profiles are given in Fig. 2(c). In this symmetric case holes are sought in the form $\psi = q(x)e^{-i\mu t}$, where the function $q(x)$ changes its sign at $x = 0$ and asymptotically approaches $\pm a_+$ at $x \to \pm\infty$. To characterize them, we use the background amplitude $a = \max|\psi| = a_+$, the redefined norm $N = \int (a_+^2 - |\psi|^2)dx$, and the integral width of the notch $w = 2[\int x^2(a_+^2 - |\psi|^2)dx / N]^{1/2}$. One interesting feature of the holes is that they exist at both positive and negative $\mu$ values, in contrast to conventional dark solitons in BEC, which exist only at $\mu \geq 0$ [69]. The width $w$ of the holes monotonically decreases with $\mu$, while the redefined norm $N$ shows a nonmonotonic behavior. Note that both quantities diverge at a specific cutoff value of the chemical potential $\mu = \mu_{\text{co}}$ [see the right vertical dashed line in Figs. 2(a) and 2(b)] that is clearly different from the critical value $\mu_{\text{cr}}$, at which the constant-amplitude waves cease to exist [see the left vertical dashed lines in Figs. 2(a) and 2(b)]. This corresponds to a considerable shape transformations caused by the LHY correction in the vicinity of $\mu_{\text{co}}$. In Fig. 2(c) one can see that at $\mu \to \mu_{\text{co}}$ the hole transforms into a gradually separating pair of out-of-phase *kinks*. We analytically the found corresponding isolated *single-kink* solution that writes $\psi = (g_{\text{LHY}} / 3\delta g)[1 + \tanh(g_{\text{LHY}} x / 3\delta g^{1/2})]e^{-i\mu t}$ and exists only at the chemical potential $\mu = \mu_{\text{co}} = -2g_{\text{LHY}}^2 / 9\delta g$ (this solution was independently found recently in [75]) that cannot be numerically continued to other values of $\mu$. Thus, holes that can be nested only in constant-amplitude waves from the upper branch exist only at $\mu \geq \mu_{\text{co}}$, and do not have a linear limit, since their amplitude remains finite at $\mu_{\text{co}}$. We did not find holes with $a_-$ asymptotics.

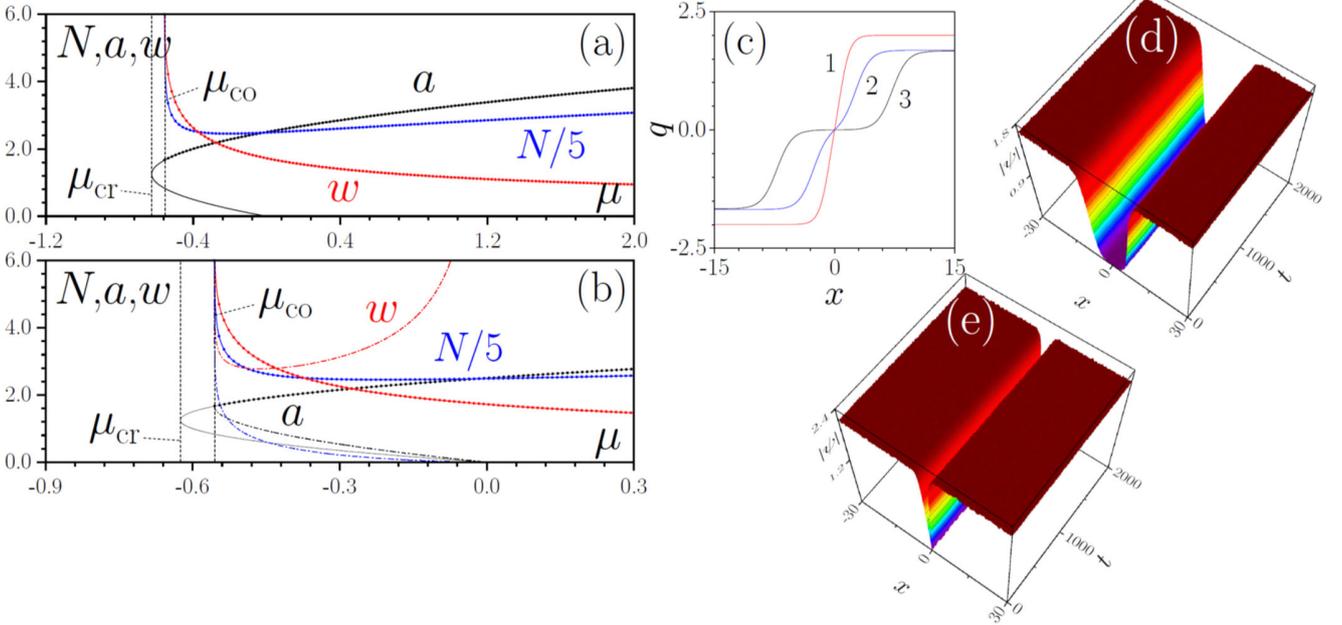

Fig. 2. (a) Redefined norm $N$, amplitude of the background $a$, and integral width of the dark notch $w$ for 1D holes versus chemical potential $\mu$ (lines with solid dots). The thin black line shows the amplitude $a_\pm$ for the two branches of the constant-amplitude solution. The vertical dashed lines denote $\mu_{\text{cr}} = -g_{\text{LHY}}^2 / 4\delta g$ and $\mu_{\text{co}} = -2g_{\text{LHY}}^2 / 9\delta g$. (b) Zoom of dependencies from panel (a) with added $N, a, w$ vs $\mu$ dependencies (dash-dotted lines) for self-bound 1D quantum droplets. (c) Examples of profiles of holes at $\mu = -0.4$ (curve 1), $\mu = -0.55$ (curve 2), and $\mu = -0.5555$ (curve 3). Stable evolution of perturbed holes with $\mu = -0.5555$ (d) and $\mu = -0.3$ (e). In all cases $\delta g = 0.4$, $g_{\text{LHY}} = 1$.

Figure 2(b) reveals the connection between the usual self-bound quantum droplets and the holes in the symmetric case. The norm $N = \int |\psi|^2 dx$, width $w = 2[\int x^2|\psi|^2 dx / N]^{1/2}$, and maximal amplitude $a = \max|\psi|$ of the droplets are plotted with dash-dotted lines. Self-bound droplets exist at $\mu_{\text{co}} \leq \mu \leq 0$. The domain of existence for droplets discussed here is consistent with the one found in Ref. [6] ($\mu_{\text{co}}$ corresponds to the chemical potential at the equilibrium density of a 1D liquid). Exactly at the same value $\mu = \mu_{\text{co}}$, at which holes transform into a pair of out-of-phase kinks, the amplitude of the droplet coincides with the amplitude of the background of the hole solution, its width and norm diverge, and the droplet transforms into a flat-top state. The latter can therefore be approximated by a kink-antikink pair (producing now a localized state rather than an extended hole region and described by the solution presented above). It is important to note that while holes can be found for $\mu_{\text{co}} \leq \mu \leq 0$ where droplet solutions are also possible, they also extend to the entire semi-infinite interval $\mu > 0$. A linear stability analysis performed for perturbed holes $\psi = [q(x) + u(x)e^{-i\delta t} + v^*(x)e^{+i\delta^* t}]e^{-i\mu t}$, where perturbations $u, v$ are $x$-dependent, within the frame of the symmetric single Eq. (1) shows that they are stable in the entire existence domain (isolated kink states are stable as well). Examples of the stable evolution of perturbed wide and narrow holes in the symmetric regime are shown in Figs. 2(d), (e).

Next, we consider a more general asymmetric situation, where we still assume equal atomic masses for species in Bose-Bose mixture, but different intra-species coupling constants. This situation occurs, in particular, in recent experiments with a mixture of $^{39}$K condensates in different hyperfine states [20]. In this case the evolution of the wavefunctions of the two different components is governed by the following system of equations [57]:

$$i\frac{\partial \psi_{1,2}}{\partial t} = -\frac{1}{2}\frac{\partial^2 \psi_{1,2}}{\partial x^2} + \frac{\partial \mathcal{E}_{1D}}{\partial n_{1,2}}\psi_{1,2}, \quad (4)$$

where $n_{1,2} = |\psi_{1,2}|^2$ are the atomic densities and the energy density including both mean field and LHY contributions is given by [6]:

$$\mathcal{E}_{1D} = \frac{1}{2}(g_{11}^{1/2}n_1 - g_{22}^{1/2}n_2)^2 + \frac{\delta g(g_{11}g_{22})^{1/2}}{(g_{11}+g_{22})^2}(g_{11}^{1/2}n_2 + g_{22}^{1/2}n_1)^2 - \frac{2}{3\pi}(g_{11}n_1 + g_{22}n_2)^{3/2}. \quad (5)$$

We chose the scaling at which $g_{11}=0.639$, $g_{22}=2.269$ and $g_{12}=-1$, which corresponds to a small $\delta g > 0$. Constant-amplitude solutions $\psi_{1,2} = a_{1,2}e^{-i\mu_{1,2}t}$ are now characterized by different amplitudes $a_{1,2}$ even for $\mu_1 = \mu_2$, but just as in the symmetric case one finds that at a fixed $\mu_2$, constant-amplitude solutions exist only above a critical value $\mu_1 = \mu_{1cr}$ of the chemical potential of the first component and that two different branches exist for $\mu_{1cr} \leq \mu_1 \leq 0$ [see the thin black and the red lines in Fig. 3(b)]. The analysis of MI of such constant-amplitude solutions, similar to that performed for the symmetric case, reveals the stability of the upper branch and the instability of the lower one. Next we considered hole states $\psi_{1,2} = q_{1,2}e^{-i\mu_{1,2}t}$ with an $x$-dependent functions $q_{1,2}$ embedded into the MI-stable background belonging to the upper branch of constant-amplitude solutions. Representative dependencies of the background amplitudes $a_{1,2}=\max|\psi_{1,2}|$, redefined total norm $N = \int[(a_1^2-|\psi_1|^2) + (a_2^2-|\psi_2|^2)]dx$, and integral width of the notch $w = 2[\int[(a_1^2-|\psi_1|^2) + (a_2^2-|\psi_2|^2)]x^2 dx / N]^{1/2}$ on chemical potential of the first component $\mu_1$ for fixed $\mu_2 = -0.2$, are presented in Figs. 3(a) and 3(b). In panel (b), for comparison, we also show the norm and width for droplets in the asymmetric case. One can see that all representative features obtained in the symmetric model are reproduced in the more general asymmetric model. This includes the transformation of holes (with different functional shapes $q_{1,2}$) into pairs of kinks [see Fig. 3(c) and 3(d)], when chemical potential of the first component approaches the value $\mu_{1co} > \mu_{1cr}$ [$\mu_{1cr}$ value corresponding to point of merger of two constant-amplitude branches is shown by the left vertical dashed line in Fig. 3(a), while $\mu_{1co}$ is shown by the right dashed line], and connection of holes at $\mu_1 = \mu_{1co}$ with the families of self-bound quantum droplets, whose amplitudes are shown by the dash-dotted lines in Fig. 3(b). Notice that because these dependencies are shown at a fixed $\mu_2$, only the second component of the quantum droplet vanishes at $\mu_1 \to 0$, while the first component becomes spatially extended with nonzero amplitude. Nearly a linear dependence of the critical values of chemical potential $\mu_{1cr}$ and $\mu_{1co}$ on $\mu_2$ is revealed by Fig. 3(e).

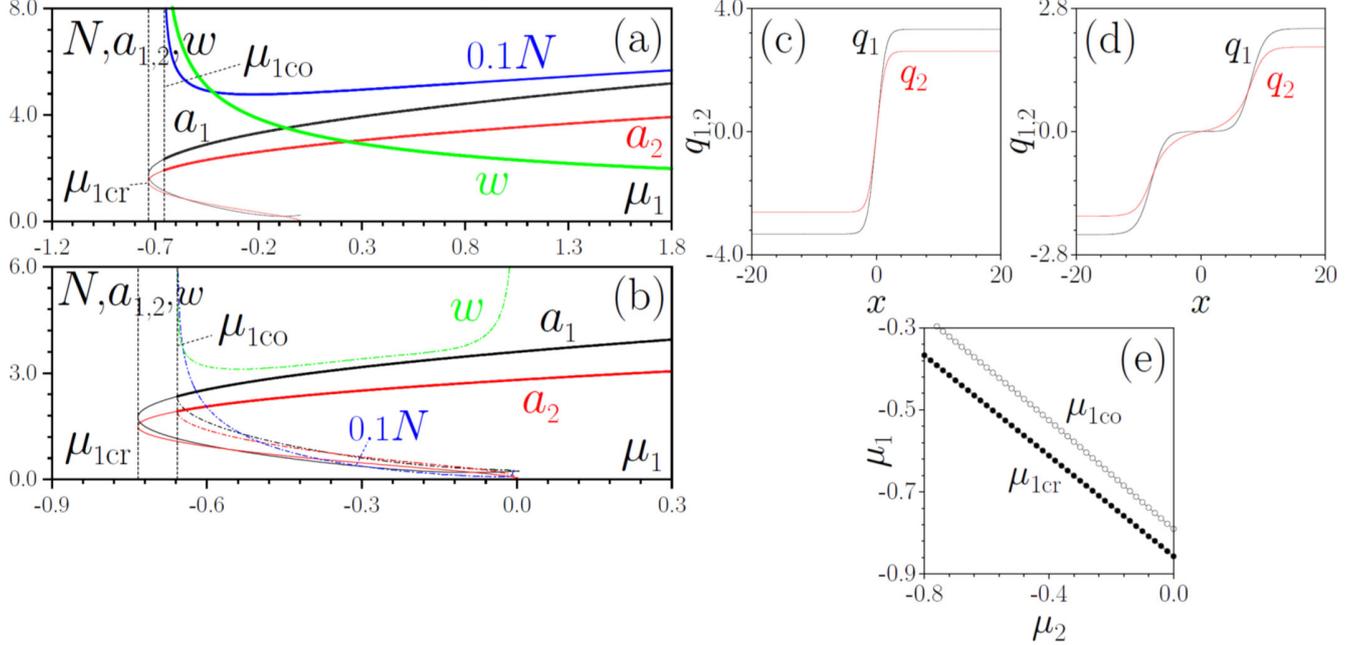

Fig. 3. (a) Redefined norm $N$, amplitudes of the background $a_{1,2}$, and integral width of the dark notch $w$ for 1D holes in the asymmetric case vs chemical potential $\mu_1$ at fixed $\mu_2 = -0.2$. The solid red and black lines show the amplitudes of the components in the constant-amplitude solution. The left vertical dashed line indicates $\mu_{1cr}$ value below which constant-amplitude solutions do not exist for selected $\mu_2$. The right vertical dashed line indicates $\mu_{1co}$ value at which holes and droplets cease to exist. (b) Zoom of $a_{1,2}(\mu_1)$ dependencies from panel (a) showing also the amplitudes, norm, and width for the droplets (dash-dotted lines). Examples of profiles of holes at $\mu_1 = -0.2$ (c) and $\mu_1 = -0.6574$ (d) at $\mu_2 = -0.2$. (e) Dependencies of $\mu_{1cr}$ and $\mu_{1co}$ on $\mu_2$. In all cases $g_{11} = 0.639$, $g_{22} = 2.269$, $g_{12} = -1$.

It turns out, however, that the stability properties of holes in asymmetric model may differ substantially from those in symmetric case. For the set of parameters considered here, self-bound quantum droplets remain completely stable, whereas we found that holes may exhibit instabilities brought by spinor nature of the model. Such instabilities manifest themselves in the breakup of the stationary holes into pairs of grey states moving in opposite directions, as shown in Figs. 4(a) and 4(b). The strength of such instability reduces with a decrease of either $\mu_1$ or $\mu_2$, as seen in the above figures, until complete stabilization occurs [Fig. 4(c)] sufficiently close to the left border of the existence domain for holes. The width of the stability domain in the asymmetric system is rather narrow; for example, for $\mu_2 = -0.2$ holes are stable only for $-0.657 < \mu_1 < -0.650$. The instability occurs only for states nested in the extended background, and is not encountered for self-bound quantum droplets.

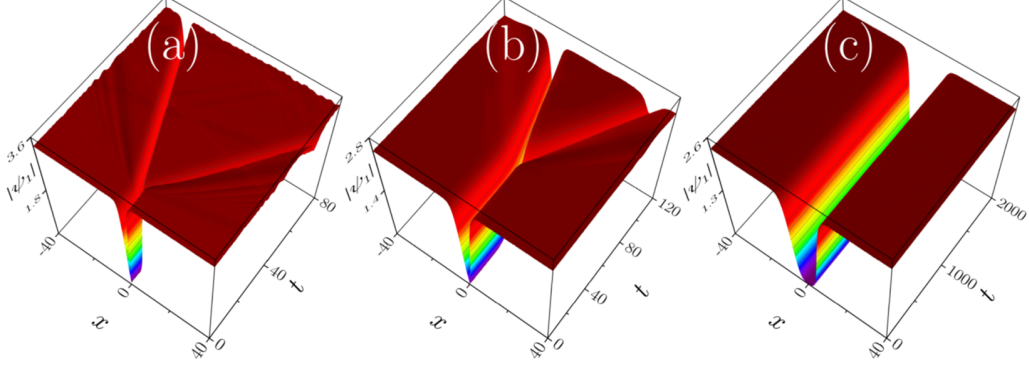

Fig. 4. Evolution dynamics of perturbed holes with $\mu_1 = -0.2$ (a), $\mu_1 = -0.62$ (b), and $\mu_1 = -0.657$ (c) in asymmetric system. In all cases $\mu_2 = -0.2$. Only the first component is shown, since dynamics of the second component is qualitatively similar. Notice different time scales. In all cases $g_{11} = 0.639$, $g_{22} = 2.269$, $g_{12} = -1$.

Next, we consider the analogs of holes in a symmetric 2D system. To this end we adopt the model describing a symmetric Bose mixture ($\psi_1 = \psi_2 = \psi$) of the form introduced in Ref. [6]:

$$i\frac{\partial \psi}{\partial t} = -\frac{1}{2}\left(\frac{\partial^2 \psi}{\partial x^2} + \frac{\partial^2 \psi}{\partial y^2}\right) + 2\alpha e^{1/2} n_0 |\psi|^2 \psi \ln(2|\psi|^2), \quad (6)$$

where $\alpha = 8\pi / \ln^2(a_{12}/a_{11})$ represents the strength of the nonlinear term whose logarithmic character stems from the LHY correction [6], $a_{12}, a_{11}$ are the corresponding inter- and intra-species scattering lengths, and the expression for $n_0$ is provided by Eq. (7) of Ref. [6]. Further we set $\alpha = 1$. First, we consider constant-amplitude solutions $\psi = ae^{-i\mu t}$ of this equation. As in the 1D case, there are two branches [see solid black lines in Fig. 5(a)]:

$$a_\pm = \left[\frac{\mu}{2\alpha W_\pm(\mu/\alpha)}\right]^{1/2}, \quad (7)$$

where $W_\pm(\mu/\alpha)$ are the upper and lower branches of the Lambert W-function. The upper branch exists for $\mu \geq \mu_{cr} = -\alpha/e$, while the lower branch exists at $\mu_{cr} \leq \mu \leq 0$. They joint at the value of the amplitude $a = (1/2e)^{1/2}$. The analysis of possible MI for perturbed constant-amplitude solutions $\psi = (a + ue^{i\mathbf{k}\mathbf{r}-i\delta t} + v^* e^{-i\mathbf{k}\mathbf{r}+i\delta^* t})e^{-i\mu t}$ gives for two branches:

$$\delta_\pm = k\left[\frac{\mu}{W_\pm(\mu/\alpha)} + \mu + \frac{k^2}{4}\right]^{1/2}. \quad (8)$$

The quantity $\mu W_+^{-1}(\mu/\alpha) + \mu$ is always positive for the entire upper branch, indicating modulational stability, but it can be negative for the lower branch, where thus MI can develop. The corresponding growth rate $\delta_{im}$ is plotted in Figs. 1(c),(d) versus $\mu$ and $k = |\mathbf{k}|$. One can see that the bandwidth of the region with MI expands with increase of the logarithmic nonlinearity strength $\alpha$ and shrinks at $\mu \to \mu_{cr}$ and at $\mu \to 0$.

Next we construct vortex-holes on modulationally stable background (upper branch) and search for them in the form $\psi(r,t) = q(r)e^{im\phi - i\mu t}$, where $m$ is the topological charge of the hole and $\phi$ is the azimuthal angle. Note that such states are sustained by the logarithmic LHY nonlinearity and on this reason they substantially differ from conventional BEC dark vortex solitons.

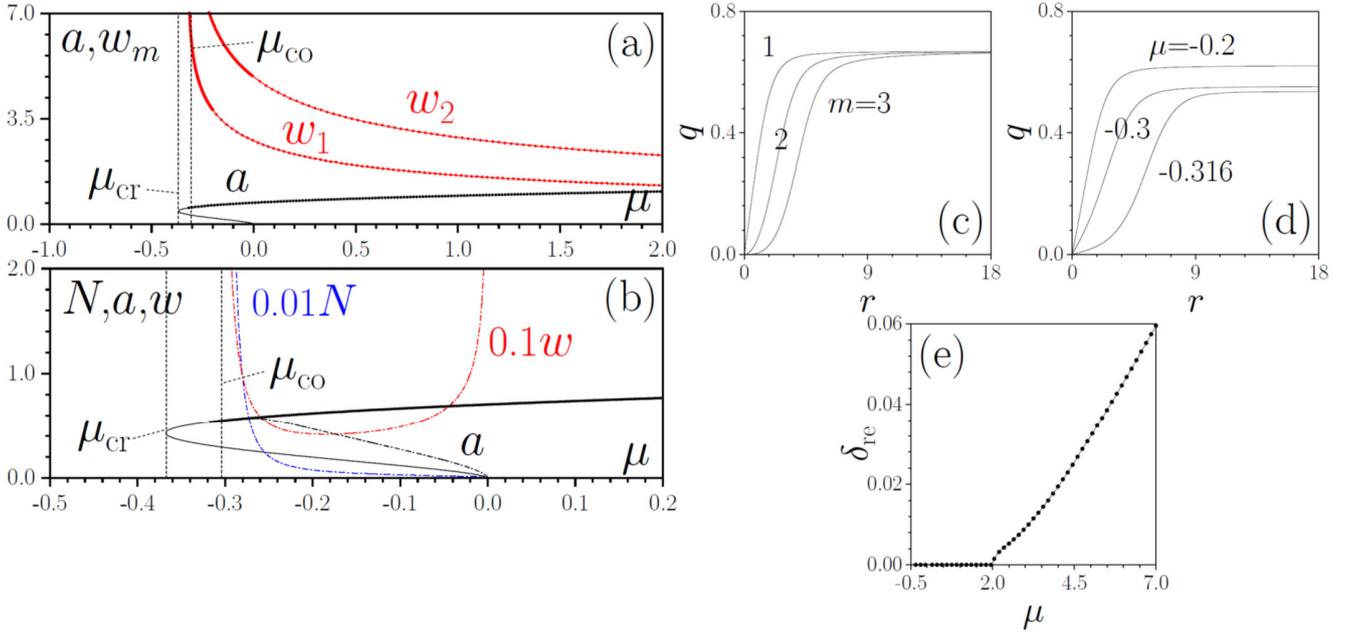

Fig. 5. (a) Amplitude of the background $a$ and width of the dark notch $w_m$ for 2D vortex-holes with topological charges $m=1$ and $m=2$ versus chemical potential $\mu$. The solid black line shows the amplitude $a_\pm$ for two branches of the constant-amplitude solution. The vertical dashed lines show $\mu_{cr} = -\alpha/e$ and cutoff value for quantum droplets $\mu_{co} = -\alpha/2e^{1/2}$. (b) Zoom of dependencies from panel (a) with added $N, a, w$ vs $\mu$ dependencies (dash-dotted lines) for 2D quantum droplets with $m=0$. Examples of profiles of vortex-holes with different topological charges at $\mu = -0.1$ (c) and different chemical potentials at $m=1$ (d). (e) Real part of perturbation growth rate vs $\mu$ for $m=1$, $n=1$. In all cases $\alpha = 1$.

The properties of the vortex-holes we found are summarized in Figs. 5(a), (b), and examples of their profiles are shown in Figs. 5(c), (d). It should be stressed that even though visually 2D vortex-holes are well-localized, their amplitude $q$ approaches the asymptotic value set by the background state $a_+$ very slowly (algebraically rather than exponentially). Therefore, even the redefined norm $N = \int 2\pi r(a_+^2 - |\psi|^2)dr$ diverges, so we used only the asymptotic amplitude and the width $w_m$ of the notch defined at the $q = a_+/2$ level to characterize the families. At fixed $\mu$, the vortex-hole with larger topological charge $m$ is broader [Fig. 5(c)]. The width monotonically decreases with the increase of the chemical potential $\mu$ [Fig. 5(a)]. We also found that the vortex-hole families cease to exist at a certain minimal value of $\mu$, at which the solution undergoes a transformation [see lower curve in Fig. 5(d)] and the tangent to the $w_m(\mu)$ dependence becomes vertical. Interestingly, this minimal value of $\mu$ is close, but does not coincide exactly with the border of existence domain for 2D quantum droplets.

In Fig. 5(b) we show with dash-dotted lines the dependence of the norm, width, and amplitude on $\mu$ for droplets with $m=0$. Their norm and width diverge at $\mu_{co} = -\alpha e^{-1/2}/2$, at peak amplitude of the state $a = 2^{-1/2} e^{-1/4}$ (flat-top regime). The corresponding $\mu_{co}$ value is shown by the right dashed line in Fig. 5(b), while the left dashed line indicates $\mu_{cr}$. The family of vortex-holes whose amplitude is shown with black dots extends slightly to the left of the $\mu = \mu_{co}$ line, i.e. as in the 1D case the domains of existence of self-bound droplets and hole states nearly coincide at $\mu \leq 0$, but the latter can be also found for all $\mu > 0$.

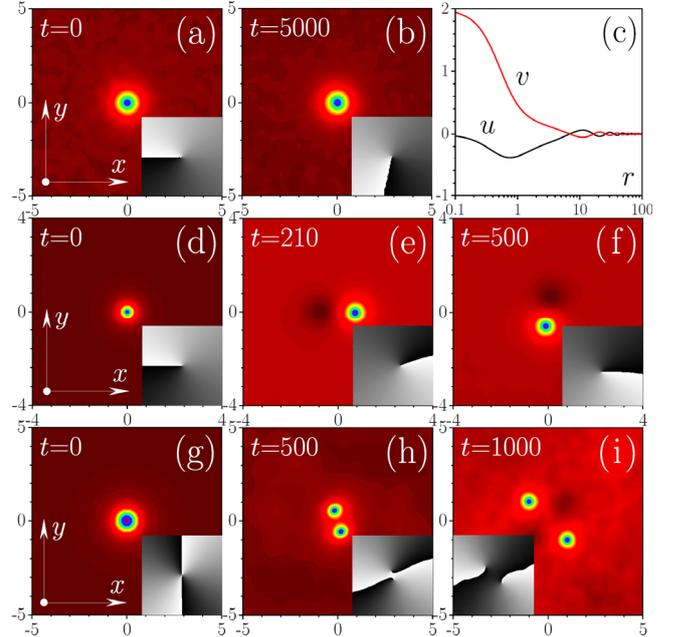

Fig. 6. (a),(b) Stable evolution of vortex-hole with $m=1$, $\mu=1$ in the presence of small noise. Main panels show $|\psi|$ in different moments of time, while insets show corresponding phase distributions. (c) Perturbation profile with $n=1$ (notice logarithmic scale for $r$ axis) that leads to instability of vortex-hole at $m=1$, $\mu=6$ illustrated in (d)-(f). (g)-(i) Decay of the unstable vortex-hole with $m=2$, $\mu=6$ stimulated by small noise. In all cases $\alpha = 1$.

To analyze the stability of the vortex-holes we substitute the perturbed solution $\psi = [q(r) + u(r)e^{in\phi+\delta t} + v^*(r)e^{-in\phi+\delta^* t}]e^{im\phi-i\mu t}$, where $n$ is the azimuthal perturbation index, into Eq. (6), linearize it and solve the corresponding eigenvalue problem. A similar linear stability analysis performed for droplets shows that they are stable in the entire domain of their existence (consistent with [34]). We found that the logarithmic nonlinearity supports stable $m=1$ holes at low and moderate $\mu$ values. An example of such stable evolution is presented in Figs. 6(a), (b). At the same time, this nonlinearity leads to nonconventional instabilities of $m=1$ holes at large $\mu$. See the dependence in Fig. 5(e) of the growth rate on $\mu$ for $n=1$ (the only perturbation that can lead to instability, in this case). The corresponding destructive perturbations are always weakly localized and feature long oscillating tails [Fig. 6(c)]. Their development leads to the formation of a bump and a hole close to it, which rotate upon evolution [Figs. 6(d)-4(f)]. Vortex-holes with $m=2$ are prone to perturbations with azimuthal indices $n=1,2$. An example of the instability development at large $\mu=6$ leading to the splitting of the initial singularity into two single-change singularities (notice the presence of bumps close to both holes at $t=1000$) is illustrated in Figs. 6(g)-6(i). At the same time, instabilities are strongly suppressed even for $m=2$ vortex-holes for $\mu \lesssim 0.5$. Even though the linear stability analysis reveals very small growth rates associated with $n=2$ perturbations for these parameters, they do not reveal themselves in dynamics and do not lead to appreciable splitting of singularities up to $t \sim 10^4$, even if considerable noise is added into the initial distributions [see the example of metastable evolution of the state with $m=2$, $\mu=0.5$ in Fig. 7].

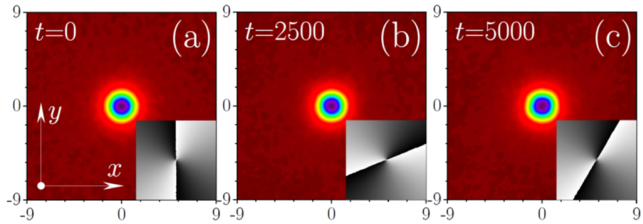

Fig. 7. Metastable evolution of vortex-hole with $m=2$, $\mu=0.5$ in the presence of small noise at $\alpha=1$.

In summary, we first have reported the existence of stable 1D holes and 2D vortex-hole states within the framework of symmetric models. Then, we have discovered instabilities of the hole states that are induced by the spinor nature of the asymmetric model. We revealed the connection between the extended states and their localized counterparts and found that they transform into kinks. In both 1D and 2D cases, the LHY contribution to the acting mean-field nonlinearity strongly impacts the shapes and properties of holes, resulting for example, in inhibited instabilities of double-charge vortex states. By and large, our results highlight the richness of the dark-like nonlinear states that are possible as a result of the competition between mean-field nonlinearities and LHY effects and, also, the remarkably different behavior predicted for systems described by scalar and spinor models. Our results may be potentially extended to 3D settings and to the case of dipolar condensates, where we anticipate the formation of more exotic states whose stability may be mediated by the LHY quantum corrections, such as vortex lines and rings.

**Acknowledgements:** Y.V.K. and L.T. acknowledge support from the Government of Spain (Severo Ochoa CEX2019-000910-S), Fundació Cellex, Fundació Mir-Puig, Generalitat de Catalunya (CERCA). M.M. acknowledges support through Grant No. PGC2018-101355-B-I00 funded by MCIN/AEI/10.13039/501100011033 and by ERDF "A way of making Europe", and by the Basque Government through Grant No. IT986-16.